# Broadband and Wide Angle Nonreciprocal Thermal Emission from Weyl Semimetal Structures

ANDREW BUTLER[1] AND CHRISTOS ARGYROPOULOS[2,*]

[1]Department of Electrical and Computer Engineering, University of Nebraska-Lincoln, Lincoln, Nebraska 68588, USA
[2] Department of Electrical Engineering, The Pennsylvania State University, University Park, PA 16803, USA
*cfa5361@psu.edu

**Abstract:** Nonreciprocal thermal emission is a cutting-edge technology that enables fundamental control over thermal radiation and has exciting applications in thermal energy harvesting. However, so far one of the foremost challenges is making nonreciprocal emission to operate over a broad wavelength range and for multiple angles. In this work, we solve this outstanding problem by proposing three different types of structures always utilizing only one Weyl semimetal (WSM) thin film combined with one or two additional dielectric or metallic layers and terminated by a metallic substrate. First a tradeoff relationship between the magnitude and bandwidth of the thermal nonreciprocity contrast is established based on the thickness of the WSM film. Then, the bandwidth broadening effect is demonstrated via the insertion of a dielectric spacer layer that can also be fine-tuned by varying its thickness. Finally, further control on the resulting strong nonreciprocal thermal radiation is demonstrated by the addition of a thin metallic layer in the proposed few layer designs. The presented composite structures work for a broad frequency range and multiple emission angles, consisting highly advantageous properties to various nonreciprocal thermal radiation applications. Moreover, the proposed designs do not require any patterning and can be experimentally realized by simple deposition fabrication methods. They are expected to aid in the creation of broadband nonreciprocal thermal emitters that can find applications in new energy harvesting devices.

## 1. Introduction

Thermal radiation is a fundamental natural phenomenon where light is spontaneously emitted from an object due to increased heating [1,2]. The photonic engineering of thermal radiation has enabled exciting and novel applications such as passive radiative cooling [3–5] and thermophotovoltaics [6,7]. Typically, thermal radiation is governed by Kirchhoff's law which equates the thermal emissivity of an object with its absorptance when it is in thermal equilibrium. However, emerging materials with asymmetric permittivity tensors can break Kirchhoff's law resulting in nonreciprocal thermal radiation [8]. Such nonreciprocal thermal emitters will have exciting applications in energy harvesting devices since they can drastically improve their efficiency by reducing energy loss to unwanted thermal radiation channels [9,10].

One of the most common ways to design nonreciprocal thermal emitters is by using magneto-optical materials that utilize an external magnetic field to create asymmetric off-diagonal components in the material's permittivity tensor [11,12]. Indium Arsenide (InAs) is a common magneto-optical material used in thermal radiation applications. Nonreciprocal absorption and emissivity have been theoretically predicted in InAs gratings when an external magnetic field is applied [8]. Another recently introduced theoretical design was able to reduce the required magnetic field strength to break reciprocity by using a silicon (Si) grating coupled to an InAs layer [13]. In terms of experiments, nonreciprocal absorption in the mid-infrared (mid-IR) has just been demonstrated by using a similar structure of a silicon carbide (SiC) grating on a InAs layer [14]. In addition, a relevant design theoretically proposed tunable nonreciprocity by introducing a graphene monolayer between a metallic grating and InAs layer [15]. Highly directional emission was also achieved by using an InAs grating embedded inside an aluminum (Al) cavity producing strong nonreciprocity at angles extremely close to the normal direction [16]. An alternative magneto-optical material is indium antimonide (InSb) that was used to achieve broadband nonreciprocal thermal emission in a recent work [17]. An unfortunate drawback of all these designs, however, is the requirement of an external magnetic field which can severely limit their realistic application, since magnets are bulky, expensive, and lossy. Finally, an alternative theoretical approach to achieve nonreciprocal thermal emission has been proposed based on modulating in time the graphene's conductivity [18,19], which consists another way to break reciprocity by applying external bias [20]. While it is an interesting idea, the rapid time-modulation of graphene properties is very challenging to be achieved experimentally. Note that most aforementioned designs are based on the diffraction mechanism due to the utilization of gratings that inherently limit the bandwidth response of nonreciprocal thermal emission.

Lately, increased attention has been dedicated to Weyl semimetals (WSMs) that are materials with inherent asymmetric permittivity tensors [21]. The intrinsic asymmetric optical constants of WSMs are an attractive option to achieve nonreciprocal radiation without the limitations stemming from external biases, such as magnets and time-modulation. Gratings composed of WSMs have been theoretically shown to produce nonreciprocal emissivity [22]. Violation of Kirchhoff's Law of thermal radiation has been shown in WSM thin films over a metallic substrate [23]. Multilayer photonic crystal designs using many WSMs layers have also been proposed to enhance nonreciprocity [24,25]. Moreover, optical Tamm states formed by multilayer WSM-based photonic crystals have been shown to enhance nonreciprocity [26,27]. One problem pervading all the aforementioned works is that the nonreciprocity is narrowband, mainly due to the used grating and photonic crystal designs and works only for a specific range of angles. In addition, complex nanopatterning or multilayer deposition methods need to be employed to achieve these types of structures [28] that are very difficult to implement in practice, especially when emerging materials such as WSM are employed that are still immature in term of fabrication. Hence, one of the foremost challenges is designing WSM-based structures that achieve broadband and omnidirectional nonreciprocal thermal emission that can be realized with as simple fabrication process as possible. One recent work theoretically proposed a Si grating over a WSM thin film and metallic substrate to achieve broad angled nonreciprocal emission in the infrared [29]. However, the results of this work are still relatively narrowband and are based on analytical modeling as opposed to the usually more accurate full-wave computational simulations.

In this work, we propose three relatively simple to fabricate designs always consisting of one WSM thin film atop a metallic substrate and combined with either one or two dielectric or metal spacer layers. First, we show that the thickness of the single WSM film affects the thermal nonreciprocity contrast bandwidth that can be extended to a very large range in the mid-IR spectrum. Next, we introduce a dielectric spacer layer and establish a tradeoff relationship between the obtained broad bandwidth and the nonreciprocity contrast magnitude. Finally, we demonstrate that the broadening effect can be tuned in mid-IR and beyond by using a thin metallic layer inserted between the WSM film and the dielectric spacer layer. All the presented designs lead to significant thermal nonreciprocity contrast values obtained over a broad wavelength range and for a wide range of emission angles. The proposed relatively omnidirectional and broadband nonreciprocity can be experimentally realized by simple deposition fabrication methods. The obtained structures are expected to have various new applications in emerging thermal energy harvesting devices.

## 2. Results

The proposed structures are always made of one WSM layer as illustrated in Fig. 1. The first design in Fig. 1(a) is a monolayer structure consisting simply of a WSM thin film of thickness $t_{WSM}$ atop a metallic substrate. This design works as an extremely simple nonreciprocal emitter. In our work, tungsten (W) is always used as the substrate metallic material that can also sustain high heating. In the second design in Fig. 1(b), a dielectric spacer layer composed of germanium (Ge) with thickness $t_{Ge}$ is introduced between the WSM layer and the substrate creating a bilayer structure. The addition of the spacer layer causes a band-broadening effect, improving the bandwidth of the nonreciprocal emission. Moreover, the choice of Ge was mainly done due to its low losses at mid-IR compared to oxide materials that exhibit high losses in this wavelength range due to the excitation of phonon-polariton resonances [30]. Finally, in the design of Fig. 1(c), an additional very thin W layer of thickness $t_W$ is added between the WSM and dielectric layer creating a trilayer structure. This metallic layer enables further tuning of the bandwidth and magnitude of the nonreciprocal response. Figure 1(d) illustrates a three-dimensional (3D) schematic of the WSM designs and depicts the dimensions of each layer.

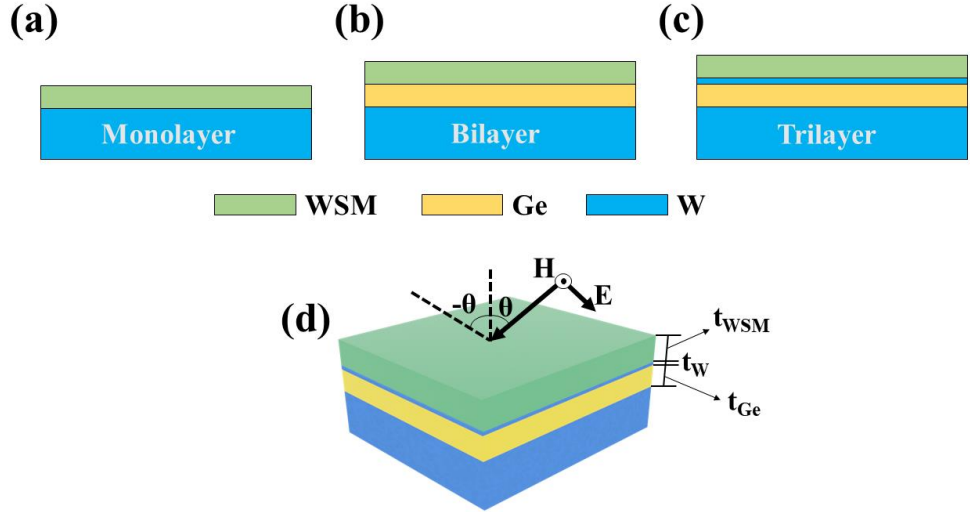

Fig. 1. (a-c) Schematics of the monolayer, bilayer, and trilayer WSM nonreciprocal thermal emitters. (d) 3D schematic of the structure with the thickness dimension of each layer also illustrated. The oblique incidence plane wave illumination is also depicted.

The reflectance, absorptance, and emissivity spectra of the proposed designs are accurately calculated based on full-wave simulations using the commercial finite element method (FEM) software COMSOL Multiphysics®. Transverse magnetic (TM) polarized plane waves are incident upon the proposed structures at an angle θ. The absorbed and reflected power is calculated for various wavelengths and incident angles to determine the absorptance and reflectance values, respectively. Periodic conditions are utilized on the side boundaries of the simulation domain. Tungsten and germanium are modelled using their experimentally derived frequency dependent complex valued optical constants [31,32], where losses are also taken into full account. The permittivity tensor of the WSM is given by the following Eq. (1) [33]:

$$\varepsilon_{WSM} = \begin{bmatrix} \varepsilon_d & j\varepsilon_a & 0 \\ -j\varepsilon_a & \varepsilon_d & 0 \\ 0 & 0 & \varepsilon_d \end{bmatrix}, \qquad (1)$$

where $\varepsilon_d$ represents the complex diagonal components of the permittivity tensor and $\varepsilon_a$ represents the off-diagonal asymmetric components that lead to nonreciprocal response. Note that all presented designs work for TM polarized light excitation since both magneto-optical materials and WSMs can have asymmetric off-diagonal permittivity components given by Eq. (1) only for this type of polarization. This poses some limitations for incoherent radiation emission but can be useful in various applications that require polarized thermal radiation, similar to the broadband directional thermal emission that has recently been experimentally demonstrated only for TM polarized light [34]. The components of the WSM permittivity tensor are plotted in Fig. 2.

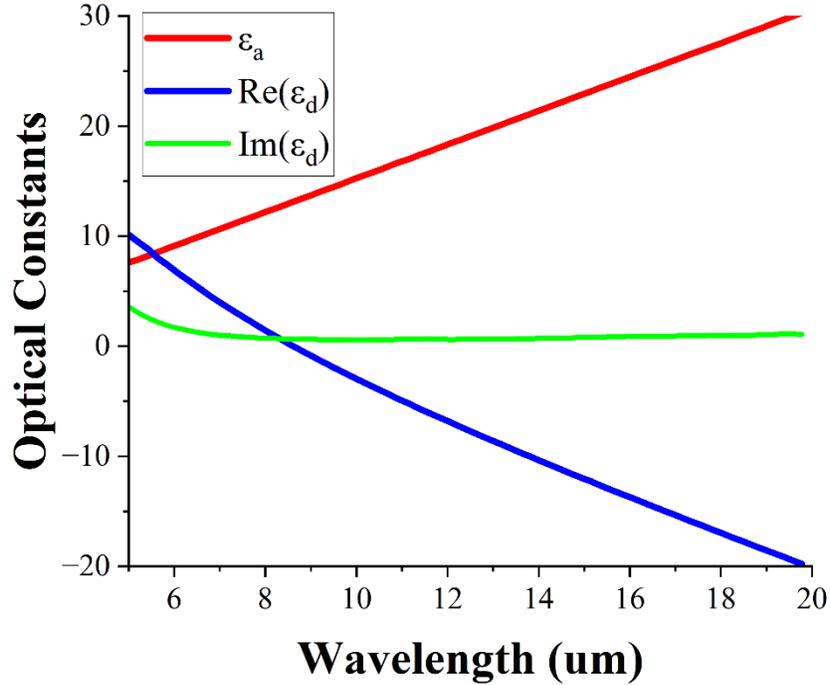

Fig. 2. Components of the WSM permittivity tensor versus wavelength. Values taken from [34].

We begin our investigation by considering the simple single layer design presented in Fig. 1(a). The reflectance as a function of wavelength and angle of incidence $R(\lambda,\theta)$ is calculated. From the reflectance, the absorptance $A(\lambda,\theta)$ and emissivity $E(\lambda,\theta)$ can be computed by using Eqs. (2)-(3) [23]:

$$A(\lambda, \theta) = 1 - R(\lambda, \theta), \qquad (2)$$
$$E(\lambda, \theta) = 1 - R(\lambda, -\theta). \qquad (3)$$

The nonreciprocity contrast metric of the structure ($\eta$) is then defined as:

$$\eta(\lambda, \theta) = |A(\lambda, \theta) - E(\lambda, \theta)|. \qquad (4)$$

This metric should be equal to zero for ordinary reciprocal structures at thermal equilibrium when computed for any wavelength and angle of incidence. However, this is not the case in the current nonreciprocal designs. In particular, Figs. 3(a) and 3(b) show the nonreciprocity contrast metric η as a function of angle and wavelength for the monolayer structure (Fig. 1(a)) with $t_{WSM}$ = 300nm and $t_{WSM}$ = 500nm, respectively. Furthermore, Figs. 3(c) and 3(d) demonstrate the quantity 1- R(λ,θ) of the same structures as a function of angle and wavelength for positive and negative incident angles. Since emissivity is calculated based on the reflectance for negative angles and absorptance is calculated with positive angles, these plots illustrate both the emissivity and absorptance. In addition, these plots should have been symmetric if the reciprocity was not broken. The nonreciprocity contrast η reaches a peak value of ~0.8 for a WSM thickness of 300nm. The peak value of η is reduced to ~0.6 when the WSM thickness is increased to 500nm with the advantage that the nonreciprocity remains high over a broader wavelength range. In both cases, the nonreciprocity metric is bifurcated into two bands. Figures 3(c) and 3(d) illustrate the reason for this bifurcation. At lower wavelengths, the absorptance (computed for θ > 0) is high, while at higher wavelengths the emissivity (calculated for θ < 0) is high. At θ = 0, the structure switches from high absorptance to high emissivity and the nonreciprocity metric becomes zero as can be seen in Figs. 3(a) and 3(b), i.e., reciprocal operation. In these designs, we limit ourselves to considering only WSM layers with thicknesses of a few hundred nanometers which is more feasible for fabrication compared to the several micron thick films used in [23].

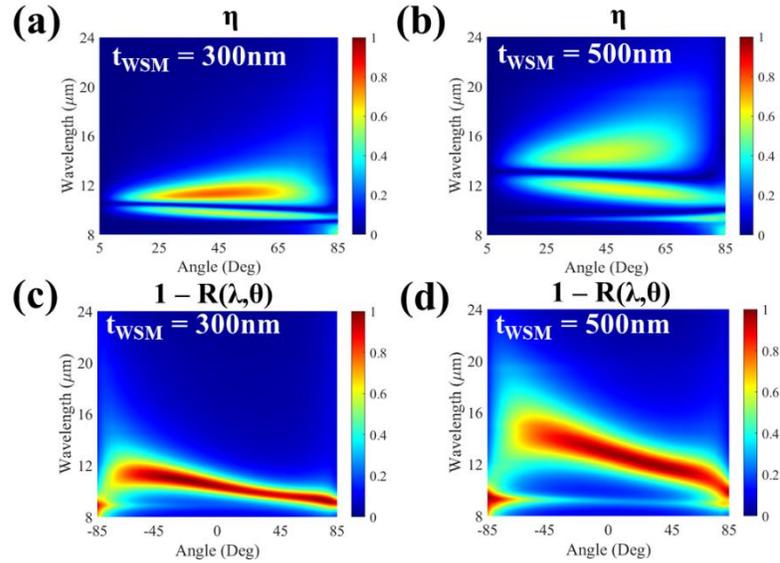

Fig. 3. (a-b) Nonreciprocity contrast metric and (c-d) 1-R(λ,θ) quantity versus angle and wavelength for the monolayer structure shown in Fig. 1(a) with WSM layer thicknesses of 300nm and 500nm, respectively.

Figure 3 clearly demonstrates that the bandwidth of the nonreciprocity increases with the thickness of the WSM layer. However, there is a small tradeoff relationship between the bandwidth and the magnitude of the nonreciprocity contrast, which led us to consider a maximum limit of 500nm thick WSM layer. Next, we study the bilayer structure illustrated in Fig. 1(b), where a thin layer of Ge is inserted between the WSM layer and the metallic substrate. This extra dielectric layer creates an additional cavity which is expected to lead to increased bandwidth in absorption or emission. Figures 4(a) and 4(b) show the nonreciprocity metric as a function of angle and wavelength for fixed $t_{Ge}$ = 150nm combined with $t_{WSM}$ = 300nm or $t_{WSM}$ = 500nm, respectively. While the magnitude of the nonreciprocity contrast is reduced, the bandwidth for both cases is substantially increased compared to the monolayer designs with results depicted before in Fig. 3. This broadening effect is further illustrated in Figs. 4(c) and 4(d), where the nonreciprocity metric for both WSM thicknesses is plotted when the incident angle is fixed to θ = 50° but for two different thicknesses of the Ge spacer layer. As the Ge thickness increases, the nonreciprocity becomes broader, spreading to a substantial part of the

IR spectrum, however the tradeoff relationship between the bandwidth and the nonreciprocity contrast magnitude turns out to be even more pronounced. This enables tuning the nonreciprocal thermal emission of the proposed structure just by changing the Ge layer thickness.

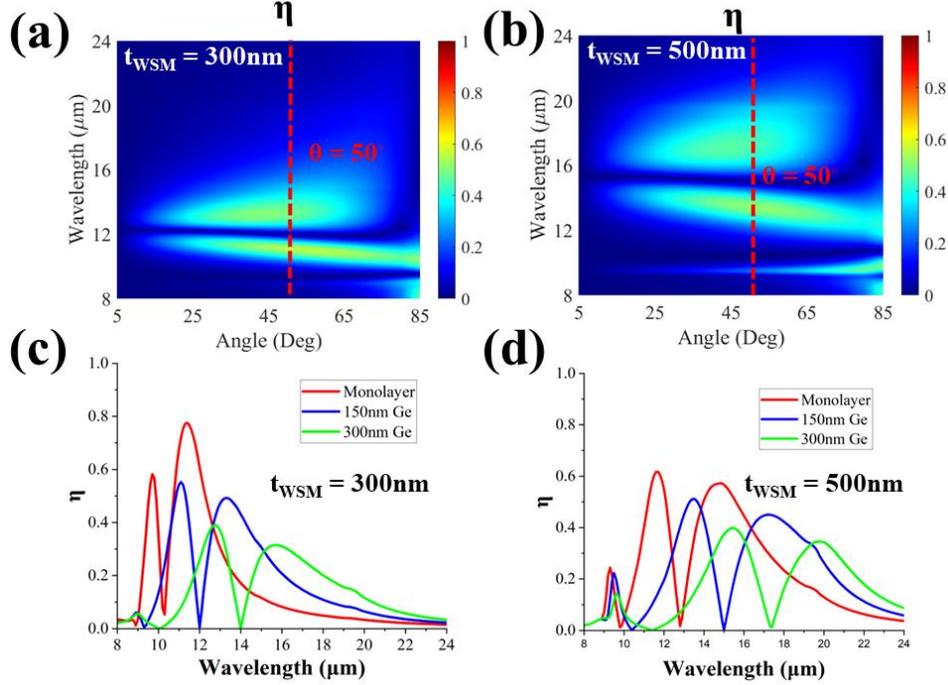

Fig. 4. (a-b) Nonreciprocity contrast metric versus angle and wavelength for the bilayer structure shown in Fig. 1(b) with fixed $t_{Ge}$ = 150nm and WSM layer thicknesses of 300nm and 500nm, respectively. (c-d) Line plots of the nonreciprocity contrast metric at fixed $\theta$ = 50° when $t_{Ge}$ is ranging from 0nm (monolayer design) to 300nm (bilayer design) and the WSM layer thicknesses are equal to 300nm and 500nm, respectively.

The cause of the broadening effect in the nonreciprocity can be further understood by examining the induced electric field distribution in the presented structures. Towards this goal, Figs. 5(a) and 5(b) depict the computed electric field enhancement induced at the monolayer structure of Fig. 1(a) for $\theta$ = -45° and $\theta$ = +45° at $\lambda$ = 11.3µm, where the nonreciprocity is the strongest. Here we consider fixed tWSM = 300nm. In both cases, the fields decay inside the WSM layer. However, the field enhancement is substantially stronger inside the WSM layer when the light is incident from the negative direction. Figures 5(c) and 5(d) demonstrate the induced field enhancement in the bilayer structure when the Ge spacer layer is introduced below the WSM layer. In this case, maximum nonreciprocity occurs at $\theta$ = ±55° and $\lambda$ = 11µm. Here, the induced field enhancement is low inside the WSM layer and negligible within the Ge spacer when the negative direction is considered. However, the induced field enhancement in the WSM is much stronger and also leaks within the Ge layer in the case of positive direction illumination. These results clearly demonstrate the nonreciprocal dynamics that lead to dissimilar field enhancement from different direction illumination.

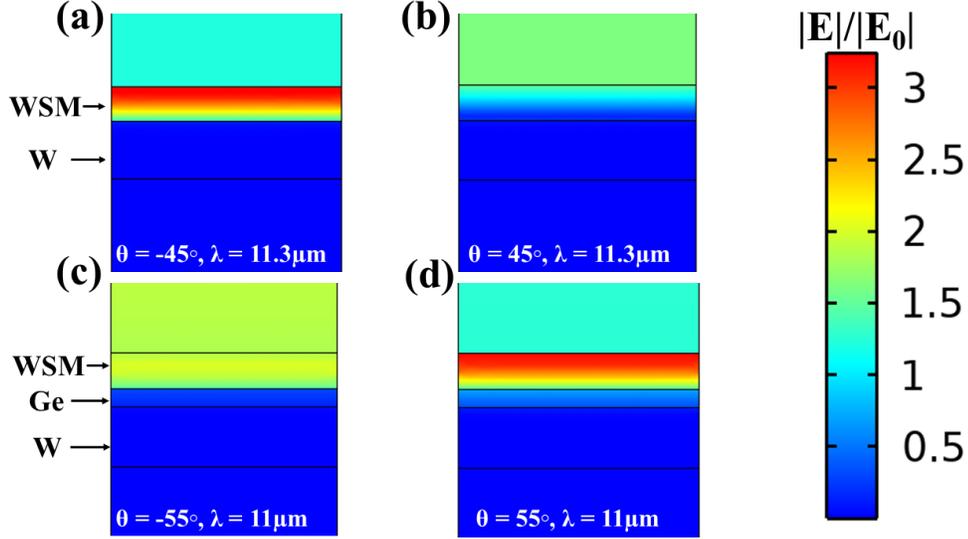

Fig. 5. (a-b) Induced electric field enhancement distribution through the monolayer structure shown in Fig. 1(a) for θ = ±45° and λ = 11.3μm. (c-d) Induced electric field enhancement distribution through the bilayer structure depicted in Fig. 1(b) for θ = ±55° and λ = 11μm.

The thermal nonreciprocity of the structures can be further tuned through the addition of an ultrathin tungsten layer, resulting in the trilayer design presented in Fig. 1(c). Figures 6(a) and 6(b) demonstrate the computed nonreciprocity contrast metric of this structure for fixed $t_W$ = 5nm, $t_{Ge}$ = 150nm and $t_{WSM}$ = 300nm or $t_{WSM}$ = 500nm, respectively. With the addition of this ultrathin metallic layer, an additional cavity is created and the nonreciprocity reaches a middle-ground between the broadband results of Fig. 4 and the high nonreciprocity contrast magnitude results of Fig. 3. Figures 6(c) and 6(d) further illustrate this tuning effect where line plots of the nonreciprocity contrast metric at θ = 50° are compared for each of the designs presented in Figs. 1(a-c) when $t_{Ge}$ is fixed to 100nm and the WSM layer thickness is equal to either 300nm or 500nm, respectively. The thickness of the metallic layer can be used to tune the response. When ultrathin, some light is prevented from entering the cavity, lessening the bandwidth broadening effect. At larger thicknesses, the resulting nonreciprocity converges towards that of the monolayer design. The final trilayer design with the ultrathin metallic layer splits the difference between the other two designs and achieves a compromise between the thermal nonreciprocity bandwidth broadening and its magnitude. Finally, it is worth commenting that the refractive index of the dielectric layer does not play a substantial role in the nonreciprocal response but low losses at mid-IR are needed to achieve the presented performance. This led us to choose Ge for the dielectric layer, which is a common semiconductor material that can be easily grown as a thin film.

Compared to other multilayer WSM designs, the nonreciprocity of our work is somewhat lower [24,26,27]. However, the nonreciprocity in these works is very narrowband and require far more layers than the designs proposed here. More complex nanoscale patterning can be used to achieve similar bandwidth with less of a decrease in nonreciprocity [28,29], however the fabrication of WSM materials is still nascent, and such patterning may be infeasible. Comparatively, our proposed structures consist relatively simple designs that can be feasible to fabricate via existing thin film deposition techniques. Towards this goal, recent experiments have begun to demonstrate the feasibility of growing single WSM thin films. More specifically, tantalum-arsenide (TaAs) is the first experimentally discovered [36,37] WSM material and recent experiments have demonstrated growth of TaAs thin films via molecular beam epitaxy [38] and pulsed laser deposition [39]. Ge thin films have been deposited on metallic substrates via electron beam evaporation [40] and thin tungsten films have been grown using atomic layer deposition (ALD) [41]. Note that the ALD technique enables highly precise control of the layer thickness, making the ultrathin 5nm tungsten layer realization also feasible [42]. Our work presents a theoretical concept that hopefully will trigger the attention of various experimental groups which are able to experimentally realize and measure such few layer structures.

The nonreciprocity broadening effect demonstrated by these results combined with efficient operation under multiple incident angles is expected to aid in the practical realization of broadband nonreciprocal thermal emitters for efficient energy harvesting applications. More specifically our work consists another step toward the development of emitters for use in nonreciprocal solar thermophotovoltaics, where maintaining high solar absorption while reducing radiated thermal emission can increase the efficiency of solar thermophotovoltaic systems to near the Landsberg limit, the theoretical maximum efficiency of solar energy harvesting (93.3%) [43]. Eliminating back thermal emission towards the sun in such systems will enable solar energy harvesting devices to approach this limit.

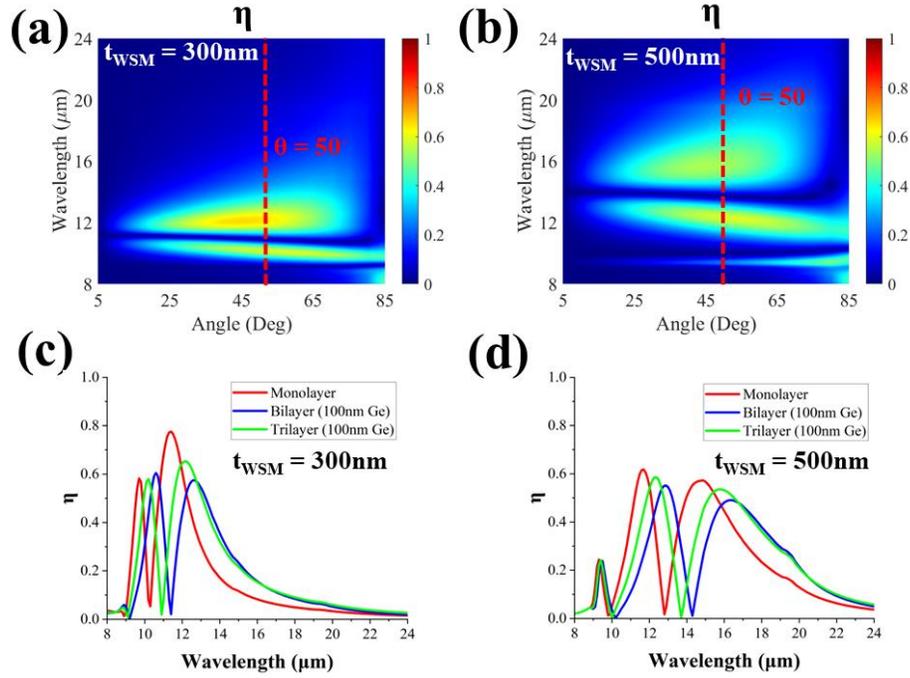

Fig. 6. (a-b) Nonreciprocity contrast metric versus angle and wavelength for the trilayer structure shown in Fig. 1(c) with fixed $t_W$ = 5nm and $t_{Ge}$ = 150nm and WSM layer thicknesses of 300nm and 500nm, respectively. (c)-(d) Line plots of the nonreciprocity contrast metric at fixed θ = 50∘ for each of the designs demonstrated in Figs. 1(a-c) when $t_{Ge}$ is fixed to 100nm and the WSM layer thicknesses are equal to 300nm and 500nm, respectively.

## 3. Conclusions

Three relatively simple to experimentally realize designs of broadband and wide angle nonreciprocal thermal emitters comprised of just one WSM layer were presented. The monolayer design demonstrated a tradeoff relationship between the nonreciprocity contrast magnitude, and its bandwidth based on the thickness of the WSM layer. The bilayer configuration showed additional frequency broadening of the nonreciprocal effect based on the inclusion of a dielectric spacer layer. Finally, the ability to tune the tradeoff relationship between nonreciprocity contrast magnitude and bandwidth was achieved via the inclusion of an additional ultrathin metallic layer leading to a trilayer nonreciprocal thermal emitter design.

The new WSM few layer structures will aid in the experimental realization of broadband and omnidirectional nonreciprocal thermal emitters. They are expected to find novel applications in emerging energy harvesting devices.

**Funding.** Nebraska Space Grant Consortium; National Science Foundation (2224456, 2212050); Office of Naval Research (N00014-19-1-2384).

**Acknowledgments.**

**Disclosures.** The authors declare no conflicts of interest.

**Data availability.** Data underlying the results presented in this paper are not publicly available at this time but may be obtained from the authors upon reasonable request.